\documentclass[preprint,showpacs,preprintnumbers,amsmath,amssymb,nofootinbib]{revtex4}
\usepackage{graphicx}
\usepackage{dcolumn}
\usepackage{bm}

\begin{document}
\title{\bf Conservation Laws for Vacuum Tetrad Gravity}

\author{Frank B. Estabrook}
\email{frank.b.estabrook@jpl.nasa.gov}
\affiliation{Jet Propulsion Laboratory, California Institute of Technology, Pasadena, CA 91109}

\date{\today}

\begin{abstract}

Ten conservation laws in useful polynomial form are derived from a Cartan form and Exterior Differential System (EDS) for the tetrad equations of vacuum general relativity.  The Noether construction of conservation laws for well posed EDSs is introduced first, and an illustration given, deriving 15 conservation laws of the free field Maxwell equations from symmetries of its EDS. The Maxwell EDS and tetrad gravity EDS have parallel structures, with their numbers of dependent variables, numbers of generating 2-forms and generating 3-forms, and Cartan character tables all in the ratio of 1 to 4.  They have 10 corresponding symmetries with the same Lorentz algebra, and 10 corresponding conservation laws.
\end{abstract}
\maketitle
\baselineskip= 14pt
\section{Introduction} In a recent paper \cite{paper1} on the mathematical structure of the WEBB tetrad partial differential equations of vacuum relativity, they were expressed as an Exterior Differential System (EDS) in a space of 44 dimensions. These equations had been formulated for numerical integration \cite{BB}\cite{ERW} and the EDS was set with variables labeled to conform with tetrad notation, as 4 to-be-independent coordinates $x^i$, 16 fields that become, in any solution, orthonormal tetrad components $^i\lambda_j$, and 24 Ricci Rotation coefficients $\Gamma_{ijk}(=-\Gamma_{ikj})$, but all treated equally ab initio.  Guided by Cartan's method of the movable frame, the EDS was given in terms of just certain combinations of basis 1-forms, viz.  $\theta_i = \eta_{ij}\mbox{ } ^j\lambda_k \mbox{ }dx^k$ and $\omega_{ij}=\Gamma_{kji}\mbox{ } ^k\lambda_l\mbox{ } dx^l$. $i, j, = 1...4.$   $ \eta_{ij}$ is a set of constants expressing the orthonormality relations and signature of the tetrad basis in a solution.  The EDS was generated by four 2-forms (torsions, $T^i =d\theta^i+\omega^i_{.j}\wedge\theta^j$) and four 3-forms for Ricci-flatness, and, for closure, the four additional 3-forms that are the exterior derivatives of the 2-forms.  A Cartan character analysis of the EDS was reported, demonstrating that it encoded a well-posed system of partial differential equations having the Cauchy properties of uniqueness, and evolution from initial conditions. Ten invariance generators of the EDS were noted. 

A Cartan 4-form $\Lambda$ was found for this EDS giving, when pulled back to a solution, a Lagrangian density $L$ for deriving the partial differential equations as the Euler-Lagrange equations of a functional variational principle. A Cartan 4-form is unique only up to the exterior derivative of an arbitrary 3-form (so-called boundary terms), but with one choice the Cartan 4-form found, evaluated on a solution, was the the Ricci scalar or Hilbert action density.  With a different choice, it evaluated to $Det|^k\lambda_l|$ times a quadratic function of the $\Gamma_{ijk}$ which we now know is found in various first order functional Lagrangian formalisms beginning with early work of Weyl \cite{W} and M{\o}ller \cite{Moller}. Functional variation of $L$ respecting the relations (from the EDS) for the ``intensity" fields $\Gamma_{ijk}$ as curls of the ``potential" fields $^k\lambda_l$ lead back to the tetrad partial differential equations.

Weyl was concerned principally with relativistic fields and unification, and there followed several long traditions of tetrad and spinor field theories with quadratic Lagrangians for generalized gravitation such as Einstein-Cartan theory and Poincar\'{e} gauge theory; cf. Pellegrini and Plebanski \cite{Pell}, Hehl \cite{Hehl}, Tupper and Phillips \cite{Tupper}, Hayashi and Nakano \cite{Naka}, Hayashi and Shirafuji \cite{Shir}, Nester and Tung \cite{Nester}, Meyer \cite{Mey}; for considerable subsequent literature cf. Itin \cite{Itin}. 

M{\o}ller in Ref. [5] saw tetrad fields as more basic than metric fields $g_{\mu \nu}$ \emph{within} classical Einstein general relativity.  From his Lagrangian, he derived an associated energy-momentum complex. The M{\o}ller energy-momentum complex for general relativity was clearly discussed by Cattaneo at GR4 in 1965 \cite{C}, by du Plessis \cite{P} and most recently, with many more citations, in a very useful review by Szabados \cite{Szabados}.  These authors emphasize that the complex associated to the M{\o}ller Lagrangian is distinct from those of Einstein/Freud, Rosen, Landau/Lifschitz, Bergmann/Komar, Goldberg and others (including earlier work by M{\o}ller.) 

In the pure vacuum case of general relativity considered in [1], the novel use of a Cartan form and EDS shows that we in fact have a classic field theory, and its ten invariance generators (or gauge transformations) enable us to go on and directly write ten polynomial conservation laws for the field equations. Four of the invariances can be understood as coming from (or describing the possibility of) arbitrary coordinate variation in a solution, and the conservation laws to which they lead, when pulled back to a solution, give a polynomial expression of the M{\o}ller complex. The other six invariances similarly correspond to arbitrary Lorentz transformation of the tetrad frame at each point of a solution, and their associated conservation laws appear to be the (distributed, nonlocal) spin of the vacuum gravity field; they all should be useful in validating the integrations of numerical relativity.

In Section II we summarize the EDS version of Noether theory, relating invariances of a Cartan form $\Lambda$ to conservation laws.  In Section III we show as an example how this works with an EDS and Cartan form for Maxwell theory, resulting in 15 conservation laws.  We point out the parallel structure to this of the tetrad EDS, Cartan form and invariance generators of Ref. \cite{paper1}, and in Sections IV and V use the latter to obtain the ten polynomial conservation laws.
\section{Noether Theory for EDS}
The generators of continuous invariance transformations of a well posed EDS have been called its \emph{isovectors}, and a systematic method for their derivation illustrated by examples, including source-free electromagnetism \cite{HE}.  An isovector is a vector field V, in the space where an EDS is given, generated by a closed exterior differential ideal $I$ ($dI \subset I$), that satisfies  
\begin{equation}
\pounds_V I \subset I
\end{equation}
Isovectors of EDSs systematize the Lie theory of invariances of systems of partial differential equations.  We note that an EDS is not unique:  sometimes some variables and generating forms can together be dropped, or it can be \emph{prolonged}, in the sense of Cartan, by introducing more (jet) variables, equivalent to higher partial derivatives, together with additional generating forms in the ideal $I$.  Or, by partially integrating, new variables and generating forms can be introduced, potentials and so-called \emph{pseudopotentials} in prolongation structures \cite{Wahl}, now often called \emph{integrable extensions} \cite{Ivey}. The well posed nature of an alternate EDS should be checked by calculation of its Cartan character table, and involutory variables.  The group of isovectors will change, and in particular more isovectors can arise, under such prolongation to new but related EDS.

A conservation law for an EDS is an n-1-form $\Psi$ (n being the number of independent variables in a solution, 4 in the EDSs treated below) \emph{not} in the $I$ generating the EDS, but whose exterior derivative $d\Psi$ \emph{is} in $I$, and hence vanishes on solutions.  By the Stokes' theorem the integral of $\Psi$ over a closed 3-boundary in a solution vanishes. Physically, this is conservation of a nonlocal quantity that is the integral of $\Psi$ over a given open three dimensional domain, for example over a 3-volume taken while holding t = constant. 

Use of a Cartan form, together with knowledge of its associated invariances (that may well be isovectors), leads to an elegant version of the Noether theorems for finding conservation laws from Lagrangians. We recall the defining property of a Cartan form $\Lambda$ for an EDS   
\begin{equation}d\Lambda \subset I \wedge I
\end{equation}
This says that the multisymplectic, or Poincare-Cartan form $d\Lambda$, when it is multiplied by--contracted with-- \emph{any} arbitrary vector, say $X$, yields a 4-form that is in $I$:
\begin{equation}X\rfloor d\Lambda \subset I
\end{equation}
In the mathematical literature \cite{BRY} a Noether vector $N$ for $d\Lambda$ is defined to satisfy  
\begin{equation}
\pounds_N d\Lambda = 0
\end{equation}
For a particular choice of boundary term for $\Lambda$ a somewhat different but natural definition of Noether vector could be
\begin{equation}
\pounds_N \Lambda \subset I
\end{equation}
Both of these definitions lead to conservation laws. A third, more restrictive, definition of Noether vector, which we shall use in the following, satisfies both of the above: 
\begin{equation}
\pounds_N \Lambda = 0
\end{equation} 
The conservation law associated to Noether vector $N$ is simply $\Psi = N\rfloor \Lambda$; if $\Psi$ is not in $I$ expanding equations (5) or (6), together with equation (3), is the required relation
\begin{equation}
d \Psi \subset I 
\end{equation}
\section{Conservation Laws for Source-Free Electromagnetism}
In Ref. \cite{paper1} it may have erroneously been implied that all previously known multisymplectic forms for Maxwell's equations were sums of terms, each of which was the product of a contact 1-form and a 4-form. One cited paper, by Robert Hermann \cite{hermann}, gave a new multisymplectic form for electromagnetism which was the product of a 2-form and a 3-form.  The resulting Cartan form and EDS, and conservation laws, provide a simple example of Noether theory that is a close parallel to those which arise for the WEBB system to be treated in Sections IV and V. For tetrad relativity the numbers of potential and intensity variables, numbers of generating 2-forms and 3-forms, number of terms in $d\Lambda$ and the Cartan characters of successive ranks are all just four times larger than those for this Maxwell EDS.

To be very explicit, the Maxwell EDS is set in a 14 dimensional space, with variables which we shall successively denote $(x^1,x^2,x^3,x^4,A_1,A_2,A_3,A_4,F_{23},F_{31},F_{12},F_{14},F_{24},F_{34})$.  This order is chosen consistently to allow cycling $1 \rightarrow 2 \rightarrow 3 \rightarrow 1$. The EDS is generated by a 2-form, its closure 3-form and a second 3-form:
\begin{eqnarray} dA_1\wedge dx^1+dA_2\wedge dx^2+dA_3\wedge dx^3+dA_4 \wedge dx^4- \nonumber \\ F_{12}
  dx^1\wedge dx^2+F_{31}
dx^1\wedge dx^3-F_{14}
  dx^1\wedge dx^4-F_{23}
  dx^2\wedge dx^3-F_{24}
   dx^2\wedge dx^4-F_{34}
   dx^3\wedge dx^4
   \end{eqnarray}
   \begin{eqnarray}
   dF_{12}\wedge dx^1\wedge dx^2+dF_{14}\wedge 
   dx^1\wedge dx^4+dF_{23}\wedge dx^2\wedge
   dx^3+ \nonumber \\ dF_{24}\wedge dx^2\wedge dx^4-dF_{31}\wedge
  dx^1\wedge dx^3+dF_{34}\wedge dx^3\wedge dx^4
   \end{eqnarray}
   \begin{eqnarray}
   dF_{12}\wedge dx^3\wedge dx^4-dF_{14}\wedge
   dx^2\wedge dx^3+dF_{23}\wedge dx^1\wedge
   dx^4+ \nonumber \\ dF_{24}\wedge dx^1\wedge dx^3+dF_{31}\wedge
   dx^2\wedge dx^4-dF_{34}\wedge dx^1\wedge dx^2
   \end{eqnarray}
   
The Cartan character table of this EDS, in the notation of Ref. \cite{paper1}, is 14(0,1,3,5)4+1;  it is well posed and the $x^i$ are involutory with $s_4 = 1$ gauge freedom.  A 4-form for radiation gauge could be added to achieve 14(0,1,3,6)4.  (With gauge similarly specialized, the WEBB characters are 44(0,4,12,24)4 \cite{paper1}.)

If the 2-form is dropped the two exact 3-forms by themselves generate a simpler well posed EDS in 10 variables for just the Maxwell field equations without potentials. The character table then is 10(0,0,2,4)4.  It has a 17 parameter isogroup \cite{HE}.  It does not however have a Cartan form.  Its simplest prolongation appears to be introduction of 4 potential fields, giving the EDS we treat here, and allowing Hermann's Cartan form.  Two of the isovectors are lost (duality rotation and field scale change) but the 15 remaining become Noether vectors having the Lie algebra of the conformal group C(3,1). They were first found by Harry Bateman in 1908 \cite{Bateman}. We give them explicitly below.  The associated 15 conservation laws were derived from them by Bessel-Hagen \cite{Bessel-Hagen}, using Noether theory.  A plethora of further, probably less interesting,  symmetries and conservation laws (beginning with Lipkin's zilch \cite{Lip}) follows from other prolongations, to jet variables and successively higher partials of the fields.  We can only refer to a sample of this literature, to recent papers with comprehensive bibliographies \cite{Kriv}\cite{Anco}\cite{Berg}.

Returning to the EDS in 14 variables, the exterior product of the two 3-forms vanishes identically, so it is clear that there is a multisymplectic form, the Hermann form, the outer product of the 2-form Equation (8) and 3-form Equation (10)
\begin{eqnarray}
d\Lambda = dA_1\wedge dF_{12}\wedge dx^1\wedge dx^3\wedge
   dx^4-dA_1\wedge dF_{14}\wedge dx^1\wedge
   dx^2\wedge dx^3+dA_1\wedge dF_{31}\wedge
  dx^1\wedge dx^2\wedge dx^4+ \nonumber \\  dA_2\wedge
 dF_{12}\wedge dx^2\wedge dx^3\wedge
  dx^4-dA_2\wedge dF_{23}\wedge dx^1\wedge
   dx^2\wedge dx^4-dA_2\wedge dF_{24}\wedge
   dx^1\wedge dx^2\wedge dx^3- \nonumber \\  dA_3\wedge
   dF_{23}\wedge dx^1\wedge dx^3\wedge
   dx^4-dA_3\wedge dF_{31}\wedge dx^2\wedge
   dx^3\wedge dx^4-dA_3\wedge dF_{34}\wedge
   dx^1\wedge dx^2\wedge dx^3- \nonumber \\  dA_4\wedge
   dF_{14}\wedge dx^2\wedge dx^3\wedge
   dx^4+dA_4\wedge dF_{24}\wedge dx^1\wedge
   dx^3\wedge dx^4-dA_4\wedge dF_{34}\wedge
   dx^1\wedge dx^2\wedge dx^4+ \nonumber \\  F_{12} dF_{12}\wedge
   dx^1\wedge dx^2\wedge dx^3\wedge dx^4-F_{14}
   dF_{14}\wedge dx^1\wedge dx^2\wedge dx^3\wedge
   dx^4+ \nonumber \\F_{23} dF_{23}\wedge dx^1\wedge dx^2\wedge
   dx^3\wedge dx^4- F_{24} dF_{24}\wedge dx^1\wedge
   dx^2\wedge dx^3\wedge dx^4+ \nonumber \\  F_{31} dF_{31}\wedge
   dx^1\wedge dx^2\wedge dx^3\wedge dx^4-F_{34}
   dF_{34}\wedge dx^1\wedge dx^2\wedge dx^3\wedge dx^4
\end{eqnarray}
Any integral of this gives a Cartan form.  Using boundary terms to eliminate bases such as $dF_{12}$  makes it easier later to pull it back into a solution to write a first order functional Lagrangian, so we take
\begin{eqnarray} \Lambda = F_{14} dA_1\wedge dx^1\wedge dx^2\wedge dx^3-F_{31}
   dA_1\wedge dx^1\wedge dx^2\wedge dx^4-F_{12}
   dA_1\wedge dx^1\wedge dx^3\wedge dx^4+\nonumber \\ F_{24} 
   dA_2\wedge dx^1\wedge dx^2\wedge dx^3+F_{23}
   dA_2\wedge dx^1\wedge dx^2\wedge dx^4-F_{12} 
   dA_2\wedge dx^2\wedge dx^3\wedge dx^4+ \nonumber \\ F_{34} 
   dA_3\wedge dx^1\wedge dx^2\wedge dx^3+F_{23}
   dA_3\wedge dx^1\wedge dx^3\wedge dx^4+F_{31}
   dA_3\wedge dx^2\wedge dx^3\wedge dx^4+ \nonumber \\ F_{34} 
   dA_4\wedge dx^1\wedge dx^2\wedge dx^4-F_{24}
   dA_4\wedge dx^1\wedge dx^3\wedge dx^4+F_{14}
   dA_4\wedge dx^2\wedge dx^3\wedge dx^4+ \nonumber \\ \frac{1}{2}   
   \left(F_{23}^2+F_{31}^2 +F_{12}^2-F_{14}^2 -F_{24}^2 -F_{34}^2 \right) dx^1\wedge dx^2\wedge dx^3\wedge dx^4
\end{eqnarray}

Pulling back this $\Lambda$ to a solution manifold, using Equation (8), yields the functional Lagrangian as usually given in terms of the fields $A_i$ (actually missing from this Lagrangian) and their partials (here in the combinations $A_{i,j}-A_{j,i}$).

The Bessel-Hagen conservation laws of vacuum electromagnetism follow directly from contraction of the vectors of the isogroup on this $\Lambda$. For conciseness we will tabulate only seven typical isovectors.  The other eight can be obtained from these by a simple cycling of the coordinate indices $1 \rightarrow 2  \rightarrow 3 \rightarrow 1$ in both the component labels and entries. The vectors denoted $L_a, L_4,a = 1,2,3$ are space-time translations, and lead to well-known energy-momentum density expressions for the vacuum electromagnetic field.  The $P_a$'s and $Q_a$'s satisfy the Lorentz Lie algebra, and give spin and cospin field densities behaving like angular momentum and center of mass motion conservation.  $R_a,R_4,R_0$ generate the extension of this to the conformal group C(3.1), and no intuitive understanding of the associated conservation laws seems to have been offered.  \begin{eqnarray}
L_1 &=& (1.0,0,0,0,0,0,0,0,0,0,0,0,0) \nonumber \\
L_4 &=& (0,0,0,1,0,0,0,0,0,0,0,0,0,0) \nonumber \\
P_1 &=& (x^4,0,0,x^1,-A_4,0,0,-A_1,F_{24},-F_{34},0,0,F_{12},-F_{31}) \nonumber \\
Q_1 &=& (0,-x^3,x^2,0,0,-A_3,A_2,0,F_{31},-F_{12},0,0,-F_{34},F_{24} \nonumber  \\
R_0 &=& (x^1,x^2,x^3,x^4,-A_1,-A_2,-A_3,-A_4,-2 F_{12},-2 F_{31},-2 F_{14},-2 F_{23},-2
   F_{24},-2 F_{34}) \nonumber \\  
R_1 &=& (\frac{1}{2} (x^{1^{2}}-x^{2{^2}}-x^{3^{2}}+x^{4^{2}}),x^1 x^2,x^1 x^3,x^1 x^4,-A_1
   x^1-A_2 x^2-A_3 x^3-A_4 x^4,\nonumber
   \\ &&
A_1 x^2-A_2 x^1,A_1 x^3-A_3 x^1,-A_4 x^1-A_1 x^4,-2
   F_{12} x^1+F_{23} x^3+F_{24} x^4,\nonumber
   \\ &&
-2 F_{31} x^1+
      F_{23} x^2-F_{34} x^4,-2 F_{14}
   x^1-F_{24} x^2-F_{34} x^3,-2 F_{23} x^1-F_{31} x^2-F_{12} x^3,\nonumber
   \\ &&
-2 F_{24} x^1+F_{14}
   x^2+F_{12} x^4,-2 F_{34} x^1+F_{14} x^3-F_{31} x^4) \nonumber \\
R_4 &=& (-x^1 x^4,-x^2 x^4,-x^3 x^4,\frac{1}{2} (-x^{1^{2}}-x^{2^{2}}-x^{3^{2}}-x^{4^{2}}),A_4
   x^1+A_1 x^4,A_4 x^2+A_2 x^4,\nonumber
   \\ &&
A_4 x^3+A_3 x^4,A_1 x^1+A_2 x^2+A_3 x^3+A_4 x^4,-F_{24}
   x^1+F_{14} x^2+2 F_{12} x^4,\nonumber
   \\ &&
F_{34} x^1-F_{14} x^3+2 F_{31} x^4, 
F_{12} x^2-F_{31}
   x^3+2 F_{14} x^4,-F_{34} x^2+F_{24} x^3+2 F_{23} x^4,\nonumber
   \\ &&
-F_{12} x^1+F_{23} x^3+2 F_{24}
   x^4,F_{31} x^1-F_{23} x^2+2 F_{34} x^4) 
 \end{eqnarray}
 
\section{Energy and Momentum Conservation in Tetrad Gravity}We now consider the $\Lambda$ of vacuum tetrad gravity,  Equation (14) of Ref. [1]:
\begin{equation}\Lambda=\omega_{is}\wedge\omega^s_{.j}\wedge\theta_k\wedge\theta_l \epsilon^{ijkl}-2\omega_{ij}\wedge\omega_{ks}\wedge\theta^s\wedge\theta_l \epsilon^{ijkl}+2\omega_{ij}\wedge T_k\wedge\theta_l \epsilon^{ijkl}
\end{equation}
$\Lambda$ has 1440 terms when written out. The coordinates $x^i$ do not explicitly appear in $\Lambda$ so it is immediate that $\pounds_{\partial/\partial  x^i}\Lambda =0$, i.e. the  $L_i$ are both isovectors and Noether vectors:
\begin{eqnarray}
L_1 = (1,0,0,0,0,....) \nonumber \\
L_2 = (0,1,0,0,0,....) \nonumber \\
L_3 = (0,0,1,0,0,....) \nonumber \\ 
L_4 = (0,0,0,1,0,....)  
\end{eqnarray}
(43 zero entries in each component list!)  We have used Mathematica to calculate the four conserved 3-forms $t_i$:
\begin{equation}
t_i = L_i \rfloor \Lambda
\end{equation}
and gone on to pull them back to a solution (using the curl relations Equation (17) of Ref. \cite{paper1}), to there expand them as polynomials on the coordinate 3-form basis:
\begin{equation}
t_p = T_p^i \mbox{ }dx^j\wedge dx^k\wedge dx^l\epsilon_{ijkl}
\end{equation}
The $T_p^i$ are evidently a non-local energy-momentum complex for gravitation, depending of course on the choice of this particular $\Lambda$, and its boundary terms, as has been explained.  $T_p^i$ has 1248 terms in each of the 12 off-diagonal components, and 1584 terms in each of the 4 diagonal components!  They are homogeneous polynomials, each term being quartic in the $^i\lambda_j$ and quadratic in the $\Gamma^i_{jk}$, and can readily be evaluated point-by-point for integration, given a numerical solution for these 40 fields. The apparent complexity of $T_p^i$ can however be greatly reduced by using summations over the tetrad indices and by introducing the transpose inverse fields $_i\lambda^j$ (cf. Ref. \cite{paper1}.  These arrays are the dual contravariant tetrad basis vectors in solutions.) Perhaps the most succinct way of writing the result is then the following:
\begin{equation}
T_p^q = Det|^i\lambda_j|\mbox{ } {_i}\lambda^q \mbox{ } \mathcal{C}^i_j \mbox{ } ^j\lambda_p
\end{equation}
where
\begin{equation}
\mathcal{C}^i_j = \mathcal{Z}^i_j + \mathcal{Y}^i_j + \mathcal{H}^i_j + \mathcal{R}^i_j
\end{equation}
and
\begin{eqnarray}
\mathcal{Z}^i_j &=& (1/2)(\Gamma^k_{.kl}\Gamma^{li}_{..j}-\Gamma^{ki}_{..j}\Gamma^ l_{.kl}) \nonumber \\ 
\mathcal{Y}^i_j &=& \Gamma^{kli}\Gamma_{lkj}-\Gamma^k_{.kj}\Gamma^{.li}_{l} \nonumber \\
\mathcal{H}^i_j &=& \Gamma^{kil}\Gamma_{jkl}-\Gamma^{k}_{.kl}\Gamma^{.il}_{j} \nonumber \\
\mathcal{R}^i_j &=& (1/2) (\Gamma^k_{.kl}\Gamma^{m.l}_{.m.}-\Gamma^{k.l}_{.m}\Gamma^{m}_{.kl})\delta^i_j 
\end{eqnarray}
Equations (18-20) are the first principal result of this paper, and express in explicit, calculable, form the nonlocal conservation laws for energy/momentum found by M{\o}ller \cite{Moller}.  M{\o}ller also sought to justify adjoining six additional equations to achieve locality;  these we now recognize as the ''Lorentz" gauge conditions of van Putten and Eardley \cite{paper1}.  Without restricting the gauge, in Section V we next find six additional nonlocal conservation laws .

\section{Six Spin Conservation Laws for Tetrad Gravity}Solutions of the tetrad equations go into neighboring solutions under arbitrary orthogonal or Lorentz transformation of the tetrads, that is, keeping the $\eta_{ij}$ unchanged.  While physically identified, these are mathematically distinct, so generated by isovectors of the EDS and leading to nontrivial conservation laws. Both $\Lambda$ and the generators of the EDS, Equation (6) of Ref. \cite{paper1}, are covariant with respect to Lorentz indices of the $^i\lambda_j$ and $\Gamma_{ijk}$ so the components of these isovectors can be written by inspection.  We will denote them $P_a$ and $Q_a$ as they satisfy the same Lorentz Lie algebra as the Maxwell invariances of Section III.  First, recall the order of variables adopted in Ref. \cite{paper1}:
\begin{eqnarray}
 x^1,x^2,x^3,x^4,^1\lambda_1,^1\lambda_2,^1\lambda_3,^1\lambda_4,^2\lambda_1,^2\lambda_2,^2\lambda_3,^2\lambda_4,^3\lambda_1,^3\lambda_2,^3\lambda_3,^3\lambda_4,^4\lambda_1,^4\lambda_2,^4\lambda_3,^4\lambda_4, \nonumber \\ \Gamma_{112},\Gamma_{113},\Gamma_{114},\Gamma_{123},\Gamma_{124},\Gamma_{134},\Gamma_{212},\Gamma_{213},\Gamma_{214},\Gamma_{223},\Gamma_{224},\Gamma_{234},\Gamma_{312},\Gamma_{313},\Gamma_{314}, \nonumber \\ \Gamma_{323},\Gamma_{324},\Gamma_{334},\Gamma_{412},\Gamma_{413},\Gamma_{414},\Gamma_{423},\Gamma_{424},\Gamma_{434} 
\end{eqnarray}
The components of two isovectors then express, respectively an infinitesimal tetrad boost in the $_1\lambda^i$ direction 
\begin{eqnarray}
P_1 & = &
(0,0,0,0,-^4\lambda_1,-^4\lambda_2,-^4\lambda_3,-^4\lambda_4,0,0,0,0,0,0,0,0,-^1\lambda_1,
-^1\lambda_2,-^1\lambda_3,-^1\lambda_4 , 
\nonumber 
\\ 
&& \Gamma_{412}-\Gamma_{124},\Gamma_{413}-\Gamma_{134},\Gamma_{414},\Gamma_{423}, 
\nonumber 
\\
&& \Gamma_{424}-\Gamma_{112},\Gamma_{434}-\Gamma_{113},-\Gamma_{224},-\Gamma_{234},0,0,-\Gamma_{212},
-\Gamma_{213},-\Gamma_{324}, 
\nonumber 
\\
&& -\Gamma_{334},0,0,-\Gamma_{312},-\Gamma_{313},\Gamma_{112}-\Gamma_{424},\Gamma_{113}-\Gamma_{434}, 
\nonumber 
\\ 
&& \Gamma_{114},\Gamma_{123},\Gamma_{124}-\Gamma_{412},\Gamma_{134}-\Gamma_{413})
\end{eqnarray}
and rotation about $_1\lambda^i$:

\begin{eqnarray}
Q_1 &=&
(0,0,0,0,0,0,0,0,^3\lambda_1,^3\lambda_2,^3\lambda_3,^3\lambda_4,
-^2\lambda_1,-^2\lambda_2,-^2\lambda_3,-^2\lambda_4,0,0,0,0, 
\nonumber 
\\
&& \Gamma_{113},-\Gamma_{112},0,0,\Gamma_{134},-\Gamma_{124},\Gamma_{213}
+\Gamma_{312},\Gamma_{313}-\Gamma_{212},\Gamma_{314},\Gamma_{323},\Gamma_{234}+ 
\nonumber 
\\
&& \Gamma_{324},\Gamma_{334}-\Gamma_{224},\Gamma_{313}-\Gamma_{212},-\Gamma_{213} 
\nonumber 
\\
&& -\Gamma_{312},-\Gamma_{214},-\Gamma_{223},\Gamma_{334}-\Gamma_{224},
-\Gamma_{234}-\Gamma_{324},\Gamma_{413},-\Gamma_{412},0,0,\Gamma_{434},-\Gamma_{424})\end{eqnarray}
and the other four isovectors follow from cyclic permutations $1 \rightarrow 2  \rightarrow 3 \rightarrow 1$.  We have in the last two equations specialized to the tetrad EDS with $\eta_{ij} = diag(+,+,+,-)$, which is of most interest, and to give correspondence with the Maxwell example.

Finally it is straightforward to calculate six conserved 3-forms by contracting these vectors on $\Lambda$.  We label them $M_{pq} = - M_{qp}$.  Each of these conserved 3-forms (without further pull back) is already in terms of the four bases $dx^i\wedge dx^j \wedge dx^k$, each of whose coefficient is a polynomial with 48 terms, third order in the $^i\lambda_j$ and linear in the $\Gamma_{ijk}$. The result is, as in Section IV, most conveniently written using the contravariant fields $_i\lambda^j$, and we use raised/lowered Lorentz indices to allow for general $\eta_{ij}$:
\begin{equation}
M_{pq} = Det|^i\lambda_j|\mbox{ }(_p\lambda^l \Gamma^t_{.tq}-_q\lambda^l \Gamma^t_{.tp}-_t\lambda^l\Gamma_{p.q}^{.t}+_t\lambda^l\Gamma_{q.p}^{.t})\mbox{ }dx^i\wedge dx^j \wedge dx^k \epsilon_{ijkl}
\end{equation}
As remarked by Szabados \cite{Szabados}, these expressions were first found by M{\o}ller as ``superpotentials" for the tetrad energy/momentum complex, Eq. (18). Like the energy/momentum conservation laws, these spin conservation laws should be of use in connection with numerical integrations.
\section{Acknowledgements}   All the forms and vectors above were inserted into programs using the AVF (algebra-valued forms) Mathematica package of H. D. Wahlquist, and all our statements as to Lie derivatives, contractions, exterior derivatives, Cartan characters, etc. explicitly verified. This research was performed at the Jet Propulsion Laboratory, California Institute of Technology, under contract with the National Aeronautics and Space Administration, partially funded through the internal Research and Technology Development program.


\begin{references}

\bibitem{paper1}F.B. Estabrook, Phys. Rev. D {\bf 71}, 044004 (2005)

\bibitem{BB}L.T. Buchman and J.M. Bardeen, Phys. Rev. D  {\bf 67}, 084017 (2003)

\bibitem{ERW} F.B. Estabrook, R.S. Robinson and H.D. Wahlquist, Class. Quantum Gravity  {\bf 14}, 1237 (1997)

\bibitem{W}H. Weyl, Z. Phys. {\bf 56}, 330-352 (1929)

\bibitem{Moller} C. M{\o}ller, Mat.-Fys. Skr. K. Danske Vid. Selsk. {\bf 1}, (10), 1 (1961)

\bibitem{Pell} C. Pellegrini and J. Plebanski, Mat.-Fys. Skr. K. Danske Vid. Selsk. {\bf 2}, (4), 1 (1963)

\bibitem{Hehl} F.W. Hehl, {\it Four Lectures on Poincar\'{e} Gauge Theory} Proc. 6th Course on Spin, Torsion and Supergravity (Erice, Italy), Eds. P.G. Bergmann and V de Sabbata (Plenum, New York, 1979)

\bibitem{Tupper} B.O.J. Tupper and G.W. Phillips, J. Phys. A: Gen. Phys. {\bf 4}, 597-610 (1971)

\bibitem{Naka} K. Hayashi and T. Nakano, Prog. Theor. Phys. {\bf 38}, 491 (1967)

\bibitem{Shir} K. Hayashi and T. Shirafuji, Phys. Rev. {\bf D19}, 3524 (1979)

\bibitem{Nester} J.M. Nester and R. S. Tung, Gen. Rel. Grav. {\bf 27} 115-9 (1995); Phys. Rev. D {\bf 60} 021501 (1999)

\bibitem{Mey} H. Meyer, Gen. Rel. Grav. {\bf 14}, 531-547 (1982)

\bibitem{Itin} Y. Itin, Class. Quantum Grav. {\bf 19}, 173-189 (2002)

\bibitem{C} C. Cattaneo, {\it Conservation Laws}, Proceedings of Fourth International Conference on GRG, Relativistic Theories of Gravitation (published for participants, Kings College, London, 1965); Ann. Inst. Henri Poincar\'{e}, {\bf IV}, 1, 1-20 (1966); extensive bibliography

\bibitem{P} J.C. du Plessis, {\it Invariance Properties of Variational Principles in General Relativity} (Math. Comm. Univ. S. Africa, Pretoria, 1968).  Discusses generalization to arbitrary, non orthonormal, tetrads.

\bibitem{Szabados} L.B. Szabados, Living Rev. Relativity {\bf 7}, 4 (2004)

\bibitem{HE}B.K. Harrison and F.B. Estabrook, J. Math. Phys. {\bf 12}, 653-666 (1971); a review of subsequent work using differential forms and Lie derivatives to find symmetries of partial differential equations is: B.K. Harrison, SIGMA {\bf 1}, Paper 001, math-ph/0510068 (2005)


\bibitem{Wahl}H.D. Wahlquist and F.B. Estabrook, J. Math. Phys. {\bf 16}, 1 (1975);  F.B. Estabrook and H.D. Wahlquist, J. Math. Phys. {\bf 17}, 1293 (1976); J.D. Finley and J.K. McIver, Acta Appl. Math., {\bf 32} 197 (1993)

\bibitem{Ivey}T.A. Ivey and J.M. Landsberg, {\it Cartan for Beginners: Differential Geometry via Moving Frames and Exterior Differential Systems}, (American Mathematical Society, Providence; Graduate Studies in Mathematics, Volume 61, 2003)

\bibitem{BRY}R. Bryant, P. Griffiths and D. Grossman {\it Exterior Differential Systems and Euler-Lagrange Partial Differential Equations} (The University of Chicago Press, Chicago, London, 2003)

\bibitem{hermann}R. Hermann, Acta Appl. Math. {\bf 12}, 35 (1988)

\bibitem{Bateman}H. Bateman, Proc. Lond. Math. Soc. {\bf 8}, 223 (1910)

\bibitem{Bessel-Hagen}E. Bessel-Hagen, Math. Annalen {\bf 84}, 258 (1921)

\bibitem{Lip}D.M. Lipkin, J. Math. Phys. {\bf 5}, 696 (1964)

\bibitem{Kriv}I.Yu. Krivskii and V.M. Simulik, Teor. Mat. Fizika, {\bf 80}, No. 2, 274 (1989); Theoretical and Math. Physics, {\bf 80} (2), 864(1989)

\bibitem{Anco} S.C. Anco and J. Pohjanpelto, Acta. Appl. Math. {\bf 69}, 285 (2001)

\bibitem{Berg} G. Bergqvist, I. Eriksson and J.M.M. Senovilla, Class. Quantum Gravity {\bf 20}, 2663 (2003)

\end{references}
\end{document}